\documentclass[11pt,preprint,tighten,flushrt]{aastex}   

\newcommand{\beq}{\begin{equation}}
\newcommand{\eeq}{\end{equation}}
\newcommand{\beqa}{\begin{eqnarray}}
\newcommand{\eeqa}{\end{eqnarray}}

\def\eq#1{equation~(\ref{#1})}
\def\lexp{\mathop{{\langle}}}
\def\rexp{\mathop{{\rangle}}}

\def\ut{{\tilde u}}

\font\BFd=cmmib10
\font\BFt=cmmib10
\font\BFs=cmmib10 scaled 700

\def\bb#1{\relax 
\ifmmode\mathchoice
{{\hbox{\BFd #1}}}{{\hbox{\BFt #1}}}
{{\hbox{\BFs #1}}}{{\hbox{\BFs #1}}}
\else \mbox{#1} \fi }

\def\r{{\bb{r}}}
\def\k{{\bb{k}}}
\def\q{{\bb{q}}}
\def\x{{\bb{x}}}
\def\y{{\bb{y}}}
\def\khat{\hat{\k}}
\def\rhobar{{\bar \rho}}
\def\deltabar{{\bar \delta}}
\def\halo{{\rm halo}}
\def\lin{{\rm lin}}
\def\nl{{\rm nl}}
\def\om{\Omega_m}
\def\ov{\Omega_\Lambda}

\begin{document}

\title{Deriving the Nonlinear Cosmological Power Spectrum and
Bispectrum from Analytic Dark Matter Halo Profiles and Mass Functions}

\author{Chung-Pei Ma}

\affil{Department of Physics and Astronomy, University of Pennsylvania, 
Philadelphia PA 19104; cpma@physics.upenn.edu}

\and

\author{J. N. Fry}

\affil{Department of Physics, University of Florida, 
Gainesville FL 32611-8440; fry@phys.ufl.edu}

\begin{abstract}
We present an analytic model for the fully nonlinear power spectrum
$P$ and bispectrum $Q$ of the cosmological mass density field.  The
model is based on physical properties of dark matter halos, with the
three main model inputs being analytic halo density profiles, halo
mass functions, and halo-halo spatial correlations, each of which has
been well studied in the literature.  We demonstrate that this new
model can reproduce the power spectrum and bispectrum computed from
cosmological simulations of both an $n=-2$ scale-free model and a
low-density cold dark matter model.  To enhance the dynamic range of
these large simulations, we use the synthetic halo replacement
technique of Ma \& Fry (2000), where the original halos with
numerically softened cores are replaced by synthetic halos of
realistic density profiles.  At high wavenumbers, our model predicts a
slope for the nonlinear power spectrum different from the often-used
fitting formulas in the literature based on the stable clustering
assumption.  Our model also predicts a three-point amplitude $Q$ that
is scale dependent, in contrast to the popular hierarchical clustering
assumption.  This model provides a rapid way to compute the mass power
spectrum and bispectrum over all length scales where the input halo
properties are valid.  It also provides a physical interpretation of
the clustering properties of matter in the universe.
\end{abstract}

\subjectheadings{cosmology : theory -- dark matter -- large-scale
structure of universe}

\section{Introduction}

Two conceptual pictures of galaxy clustering have been examined in the
literature, the continuous hierarchical clustering model and the
power-law cluster model (Peebles 1980, \S 61).  In the hierarchical
clustering model, which has emerged as the accepted model over the
past two decades, galaxy clustering is characterized by power-law
correlation functions: the $N$-point correlation function $\xi_N$
scales with configuration size as $\xi_N \propto r^{-\gamma_N} \propto
\xi_2^{(N-1)}$, where $ \gamma_N = (N-1) \gamma $ and the two-point
correlation function goes as $\xi_2=\xi\propto r^{-\gamma} $.  The
hierarchical model is motivated by the observed power-law behavior
$\gamma\approx 1.8$ of galaxy correlations (Groth \& Peebles 1977; Fry
\& Peebles 1978), with a theoretical basis in a self-similar,
scale-invariant solution to the equations of motion (Davis \& Peebles
1977).

The alternative power-law cluster model has an even longer history
(Neyman \& Scott 1952; Peebles 1974, 1980; McClelland \& Silk 1977;
Scherrer \& Bertschinger 1991; Sheth \& Jain 1997; Valageas 1998; Yano
\& Gouda 1999).  In this model, galaxies are placed in spherical
clumps that are assumed to follow a power-law density profile $\rho(r)
\propto r^{-\epsilon}$, with the centers of the clumps distributed
randomly.  The resulting two-point correlation function is also a
power law with a logarithmic slope $\gamma = 2\epsilon - 3$.  While it
is possible to reproduce the observed two-point function by an
appropriate choice of the power index $\epsilon = (3+\gamma)/2 \approx
2.4$, Peebles and Groth (1975) pointed out that this model produces a
three-point function that is too steep to be consistent with
observations in the Zwicky and Lick catalogs.

In an earlier paper (Ma \& Fry 2000a), we have shown that in the
nonlinear regime, the three-point correlation function $\zeta=\xi_3$
of the cosmological mass density field does not exactly follow the
prediction $\zeta\propto \xi^2$ of the hierarchical clustering model.
These conclusions are drawn from study of high resolution numerical
simulations of a cold dark matter (CDM) model with cosmological
constant and of a model with scale-free initial conditions $P(k)\sim
k^n$ with $n=-2$.  In experiments replacing simulation dark matter
halos with power-law density profiles, $\rho(r)\sim r^{-\epsilon}$, we
have demonstrated that the behavior of the correlation functions in
the nonlinear regime are determined by the halo profiles, but that it
is not possible to match both the two- and three-point correlations
with a single slope $\epsilon$.  These results differ from the
predictions of both of these two conceptual models.

In this paper, we expand our previous study of the nonlinear two- and
three-point correlation functions by investigating a new prescription
that takes into account the non-power-law profiles of halos, the
distribution of halo masses, and the spatial correlations of halo
centers.  Each of these ingredients has been well studied in the
literature.  We find that this halo model provides a good description
of the two- and three-point correlation functions in both the $n=-2$
and CDM simulations over the entire range of scales from the weak
clustering, perturbative regime on large length scales, to the
strongly nonlinear regime on small length scales.  Our result is
approximately hierarchical over an intermediate range of scales, thus
uniting the two pictures.  An independent recent study by Seljak
(2000), which appeared during completion of this work, has also
examined the two-point power spectrum in a similar construction and
has found that this type of approach can reproduce the power spectrum
in the CDM model.  The analytic model proposed here can be used to
compute the two- and three-point correlation functions and their
Fourier transforms, the power spectrum and bispectrum, over any range
of scale where the input halo properties are valid.  

In a subsequent paper (Ma \& Fry 2000c), we study the predictions of
this analytic halo model for the asymptotic nonlinear behavior of the
$N$-point correlation functions and the pairwise velocities and
examine the conditions required for stable clustering.

The outline of this paper is as follows.  In \S 2 we describe the
three input ingredients of the model: halo density profiles, halo mass
functions, and halo-halo correlations.  In \S 3 we assemble these
ingredients and construct analytic expressions for the two-point
correlation function $\xi(r)$ and the power spectrum $P(k)$.  In \S 4
we do the same for the three-point correlation function
$\zeta(r_1,r_2,r_3)$ and its Fourier transform, the bispectrum
$B(k_1,k_2,k_3)$.  In \S 5 we test the validity of this new model by
comparing its predictions with results from numerical simulations of
an $n=-2$ scale free model and a low-density CDM model with a
cosmological constant ($\Lambda$CDM).  We also present results of the
synthetic halo replacement technique used to enhance the numerical
resolution.  In \S 6 we discuss further the physical meanings and
implications of the model.  In particular, we elaborate on two
important implications of this model: deviations from the common
assumptions of stable clustering and hierarchical clustering.
Section~7 is a summary.

\section{Model Ingredients}

\subsection{Halo Mass Density Profile}

It has been suggested recently that the mass density profiles of cold 
dark matter halos have a roughly universal shape, generally independent 
of cosmological parameters (Navarro, Frenk, \& White 1996, 1997)
\begin{equation}
    {\rho(r) \over \bar\rho} = \deltabar\, u(r/R_s) \,,
\label{shape}
\end{equation}
where $\bar\delta$ is a dimensionless density amplitude, $R_s$ is a
characteristic radius, and $\bar\rho$ is the mean background density.
We consider two functional forms for the density profiles
\begin{eqnarray}
   u_{I} (x)&=&{1\over x^p (1+x)^{3-p}} \,,\nonumber\\
   u_{II}(x)&=&{1\over x^p (1+x^{3-p})} \,.
\label{u}
\end{eqnarray}
Both forms have asymptotic behaviors $x^{-p}$ at small $x$ and
$x^{-3}$ at large $x$, but they differ in the transition region.  The
first form $u_I(x)$ with $p=1$ is found to provide a good fit to
simulation halos by Navarro et al. (1996, 1997), whereas the second
form $u_{II}(x)$ with a steeper inner slope $p = 3/2$ is favored by
Moore et al. (1998, 1999).  Some independent simulations have produced
halos that are well fit by the shallower $p=1$ inner slope (e.g.,
Hernquist 1990; Dubinski \& Carlberg 1991; Huss, Jain, \& Steinmetz
1999), and others the steeper $p>1$ slope (e.g., Fukushige and Makino
1997).  Jing \& Suto (2000) have recently reported a mass-dependent
inner slope, with $p\approx 1.5$ for galactic-mass halos and $p\approx
1$ for cluster-mass halos.  Many of these authors find that the outer
profile scales as $r^{-3}$, but steeper outer profiles have also been
suggested (Hernquist 1990; Dubinski \& Carlberg 1991).  Given these
uncertainties, we will consider in this paper both types of profiles
in \eq{u}.

The parameters $R_s$ and $\bar\delta$ in equation~(\ref{shape}) are
generally functions of the halo mass $M$.  A concentration parameter, 
\begin{equation}
	c = {R_{200}\over R_s} \,,
\end{equation}
can be used to quantify the central density of a halo (Navarro et
al. 1997), where $R_{200}$ is the radius within which the average
density is 200 times the mean density of the universe.  Using
$M=800\pi\,\bar\rho\,R_{200}^3/3$, we can relate $R_s$ and
$\bar\delta$ to $M$ and $c$, where the scale radius $R_s$ is
\begin{equation} 
   R_s = {1\over c} \left( 3M\over 800\pi \bar\rho \right) ^{1/3} =
   {1.63\times 10^{-5} \over \Omega_m^{1/3}\, c} 
   \left( {M\over h^{-1} M_\odot}\right)^{1/3} \,h^{-1}{\rm Mpc}\,,
\label{Rs}
\end{equation} 
and the density amplitude $\deltabar$ is 
\begin{eqnarray}
   \bar\delta_{I} &=& {200\,c^3\over 3 [\ln(1+c) - c/(1+c)]}\,,
	\qquad p=1\,,	\nonumber\\
   \bar\delta_{II} &=& {100\,c^3\over \ln(1+c^{3/2})}\,,
   \qquad p={3\over 2} \,.
\label{delbar}
\end{eqnarray}
Typical values of $c$ are in the range of a few to ten for type I and
perhaps a factor of three smaller for type II.  There is a weak
dependence on mass, such that less massive halos have a larger central
density (e.g., Cole \& Lacey 1996; Tormen, Bouchet, \& White 1997;
Navarro et al. 1996, 1997; Jing \& Suto 2000).  This is understood in
general terms as reflecting the mean density at the redshift $z_f$
when the halo initially collapsed, $\deltabar \sim (1+z_f)^3$.  For
$\Omega=1$ this is $c \sim \sigma(M)$, or $c \propto M^{-(3+n)/6}$ in
a scale-free model.

\subsection{Halo Mass Function}

The number density of halos with mass $M$ within a logarithmic
interval is often approximated by the prescription of Press \&
Schechter (1974),
\begin{equation} 
  {dn\over d\ln M} = \sqrt{2\over\pi}\,{d\ln \sigma^{-1}\over d\ln M} 
  {\bar\rho \over M}\,\nu\, e^{-\nu^2/2} \,, \qquad 
  \nu={\delta_c \over \sigma(M)}\,,
\label{PS}
\end{equation}
where $\delta_c$ is a parameter characterizing the linear overdensity
at the onset of gravitational collapse, and $\sigma$ is the linear rms
mass fluctuations in spheres of radius $R$
\begin{equation}
   \sigma^2(M)=\int_0^\infty  {4\pi k^2 dk \over (2\pi)^3}
   \, P(k) \, W^2(kR)\,,
\label{sigma}
\end{equation} 
where $W(x)= 3(\sin x - x \cos x)/x^3 $ is the Fourier transform of a
real-space tophat window function.  The mass $M$ is related to $R$ by
$M=4\pi\bar\rho R^3/3$.  For scale free models with a power law
initial power spectrum $P\propto k^n$, this is $\sigma =
(M/M_*)^{-(3+n)/6}$.  The parameter $M_*$ characterizes the mass scale
at the onset of nonlinearity, $\sigma(M_*)=1$, and is related to the
nonlinear wavenumber $k_\nl$ (defined as ]$\int_0^{k_\nl} 4\pi k^2 dk
\, P(k) / (2\pi)^3 =1$) by
\begin{equation}
    M_\ast = {4\pi\rhobar \over 3} {B(n)\over k^3_\nl} 
    = {4\pi\rhobar \over 3} R_*^3 \,,
\end{equation}
where 
\begin{eqnarray}
    \qquad B(n) &=& (k_\nl R_*)^3 = 
    \left[(n+3)\int_0^\infty dx\,x^{n+2}\,W^2(x) 
	 \right]^{3/(n+3)} \,, \nonumber \\
  B^{(3+n)/3} &=& 
   \sin\left[(n+2){\pi \over 2}\right] \, 
   \Gamma(n+2) \, { 9 \, (2^{-n}) (3+n) \over (-n)(1-n)(3-n)} 
\end{eqnarray}
(defined for $-3\le n < 1$).  Various modifications to the
Press-Schechter mass function have been suggested (e.g., Sheth \&
Tormen 1999; Lee \& Shandarin 1999; Jenkins et al. 2000) to improve
the accuracy of the original formula.

\subsection{Halo-Halo Correlations}

Dark matter halos do not cluster in the same way as the mass density
field.  On large scales, a bias parameter $b$ is typically used to
quantify this difference.  Let $\xi_\halo(r;M,M')$ be the two-point
correlation function of halos with masses $M$ and $M'$, $\xi_\lin(r)$
be the linear correlation function for the mass density field, and
$P_\halo$ and $P_\lin$ be the corresponding power spectra.  On large
length scales, we assume a linear bias and write
\begin{eqnarray}
   \xi_\halo(r;M,M') &=& b(M)\,b(M')\,\xi_\lin(r)\,, \nonumber\\
    P_\halo(k;M,M') &=& b(M)\,b(M')\,P_\lin(k)\,.
\label{Pbias}
\end{eqnarray}
Based on the peak and the Press-Schechter formalism, Mo \& White
(1996) developed a model for the linear bias $b(M)$, which is later
modified by Jing (1998) to be
\begin{equation}
   b(M) = \left(1 + {\nu^2-1 \over \delta_c}\right)\,
	\left( {1\over 2\nu^4} +1 \right)^{0.06-0.02\,n}\,, \quad
	\nu={\delta_c \over \sigma(M)} \,.
\label{bM}
\end{equation} 
The original formula for $b(M)$ by Mo \& White includes only the first
factor above; the second factor, dependent on the primordial spectral
index $n$, is obtained empirically for an improved fit to simulation
results at the lower mass end (Jing 1998).  In this bias model, $b(M)$
is below unity for $M\la M_*$ (where $\sigma(M_*)=1$) and reaches
$\sim 0.5$ for $M\la 0.01 M_*$.  Small dark matter halos are therefore
anti-biased relative to the mass density.  For $M\ga M_*$, $b(M)$
increases monotonically with the halo mass and reaches $b \sim 10$ at
$M\sim 100\,M_*$.  Nonlinear effects on the bias have been studied
(Kravtsov \& Klypin 1999 and references therein), but they are
unimportant in our model because the halo-halo correlation terms
contribute significantly only on large length scales in the linear
regime (see \S 3 and 4).

Similarly, we use higher order bias parameters to relate the
higher-order correlation functions for halos and mass density.  In
this paper we examine the three-point correlation function
$\zeta(r_1,r_2,r_3)$ and its Fourier transform, the bispectrum
$B(k_1,k_2,k_3)$ (see \S 4 for a more detailed discussion).  On large
length scales where the amplitude of $\delta$ is small, perturbation
theory can be used to relate the lowest order contribution to the
bispectrum of the mass density to the linear power spectrum $P_\lin$
(Fry 1984):
\begin{eqnarray}
        B^{(0)}(k_1,k_2,k_3) &=& F_{12} \,P_\lin(k_1) P_\lin(k_2) + 
        F_{23}\, P_\lin(k_2) P_\lin(k_3) + 
        F_{31}\, P_\lin(k_3) P_\lin(k_1)\,, \nonumber\\
	F_{ij} &=& {10\over 7} + (k_i/k_j + k_j/k_i)\,
       (\khat_i\cdot\khat_j) + {4 \over 7} (\khat_i \cdot \khat_j)^2\,.
\label{Btree}
\end{eqnarray}
Using this perturbative result and the results of Mo, Jing, \& White
(1997), we can write the halo bispectrum as
\begin{eqnarray}
   B_{\halo}(k_1,k_2,k_3;M,M',M'') &=&
     \left[ b(M) b(M') b(M'')\,F_{12} + b(M) b(M') b_2(M'') \right]\,
	P_\lin(k_1)\,P_\lin(k_2) \nonumber\\
   &+& \left[ b(M) b(M') b(M'')\,F_{23}+ b(M) b_2(M') b(M'')\right]\,
	P_\lin(k_2)\,P_\lin(k_3) \nonumber \\
   &+& \left[ b(M) b(M') b(M'')\,F_{31}+ b_2(M) b(M') b(M'')\right]\,
	P_\lin(k_3)\,P_\lin(k_1) \,,\nonumber\\
   && 
\label{Bbias}
\end{eqnarray}
where $b(M)$ is given by \eq{bM}, and the quadratic bias parameter
$b_2(M)$ is
\begin{equation}
   b_2(M) = {8 \over 21} {(\nu^2 - 1) \over \delta_c} + 
   \left( {\nu \over \delta_c} \right)^2 (\nu^2 - 3) \,.
\end{equation}
For the special equilateral case of $k_1=k_2=k_3=k$, 
\eq{Bbias} simplifies to
\begin{eqnarray}
   B^{\rm eq}_{\halo}(k;M,M',M'') &=&
    \left[{12 \over 7}\, b(M) b(M') b(M'') + b(M) b(M') b_2(M'') \right.
	\nonumber\\
   &+& \left. b(M) b_2(M') b(M'') + b_2(M) b(M') b(M'') \right]
	\,P_\lin^2(k) \,.
\label{Bbias2}
\end{eqnarray}
In practice, the terms involving $b_2(M)$ in equations~(\ref{Bbias})
and (\ref{Bbias2}) make only a small net contribution.  For
simplicity, we will therefore not include this term in the subsequent
derivations and calculations.

\section{Two-Point Statistics: $\xi(r)$ and $P(k)$}

We now construct our analytic halo model for the two-point correlation
function $\xi(r)$ and the power spectrum $P(k)$.  The two-point
correlation function of the cosmological mass density field
$\delta=\delta\rho/\bar\rho$ is
\begin{equation}	
	\xi(\r) = \langle \delta(\x)\,\delta(\x+\r) \rangle \,.
\end{equation}
The Fourier transform of $\xi(r)$ is the mass power spectrum
$P(k)=\int d^3r\,e^{-i\k\cdot\r}\,\xi(r)$, which is related
to the density field in $k$-space by $\lexp \delta(\k_1) \delta(\k_2)
\rexp = P(k_1)\,(2\pi)^3 \delta_D (\k_1 + \k_2)\,$, 
where $\delta_D$ is the Dirac delta-function.

The two-point correlation function measures the excess probability
above the Poisson distribution of finding a pair of objects with
separation $r$ (Peebles 1980).  The objects can be taken to be dark
matter particles, most of which cluster gravitationally in the form of
dark matter halos.  One should therefore be able to express $\xi$ for
the density field in terms of properties of dark matter halos.  In
this picture, we can write the contributions to $\xi$ as two separate
terms, one from particle pairs in the same halo, and the other from
pairs that reside in two different halos.  In realistic cosmological
models, dark matter halos exhibit a spectrum of masses that can be
characterized by a distribution function $dn/dM$, and the halo centers
are spatially correlated.  Taking these factors into consideration, we
can write the two-point correlation function for $\delta$ in terms of
the halo density profile $u(x)$, halo mass function $dn/dM$, and
halo-halo correlation function $\xi_\halo $ discussed in \S 2.  We
write
\begin{equation}
   \xi(r) = \xi_{1h}(r) + \xi_{2h}(r) \,,
\end{equation}
where the subscripts ``$1h$'' and ``$2h$'' denote contributions from
particle pairs in ``1-halo'' and ``2-halos'', respectively, and
\begin{eqnarray}
  \xi_{1h}(r) &=& \int d^3 r' \int dM\,{dn\over dM}\,
	\deltabar^2\, u(r'/R_s)\, u(|\r'+\r|/R_s) 
	  \nonumber\\  \noalign{\smallskip}
  \xi_{2h}(r) &=& \int d^3 r'\, d^3 r'' 
      \int dM' \, {dn\over dM'}\,\deltabar'\,u(r'/R'_s) 
      \int dM'' \, {dn\over dM''}\,\deltabar''\, u(r''/R''_s) 
      \, \xi_\halo(|\r'-\r''+\r|) \nonumber\\
    &=& \int d^3 r'\, d^3 r'' 
    \left[\int dM \, {dn\over dM}\,\deltabar\, u(r'/R_s)\,b(M) \right]
    \left[\int dM \, {dn\over dM}\,\deltabar\, u(r''/R_s)\,b(M) \right]
	\nonumber\\
    && \times\,\, \xi_\lin(|\r'-\r''+\r|) \,.
\label{xi2}
\end{eqnarray}
These expressions arise from averaging over displacements $\r'$,
$\r''$ of halo centers from the particle positions $\r_1$, $\r_2$,
where $r=|\r_1-\r_2|$.  In the last expression above, we have used the
bias model of \eq{Pbias} to relate the halo-halo correlation function
$\xi_\halo$ to the linear correlation function $\xi_\lin$ of the mass
density field.

As we will show in \S 5, the dominant contribution to the two-point
correlation function in the nonlinear regime on small length scales is
from the first, 1-halo term $\xi_{1h}$ for particle pairs that reside
in the same halos.  This makes intuitive sense, because closely spaced
particle pairs are most likely to be found in the same halo.  This
term is determined by the convolution of the dimensionless density
profile with itself,
\begin{equation} 
   \lambda(x) = \int d^3y \, u(y)\, u(|\x +\y|) \,.
\label{lamb}
\end{equation}
For many forms of $u(x)$, the angular integration in this equation is
analytic, and $\lambda$ can be reduced to a simple one-dimensional
integral over $y$.  For some special cases, $\lambda$ can even be
reduced to an analytic expression.  We leave the detailed results for
$\lambda$ to the Appendix.

In $k$-space, the convolutions in \eq{xi2} for $\xi(r)$ become simple
products.  Using $\ut(q)$ to denote the Fourier transform of $u(x)$,
where $\ut(q) =\int d^3 x\, u(x)\,e^{-i\q\cdot\x} $, we can readily
transform \eq{xi2} into expressions for the mass power spectrum:
\begin{equation} 
   P(k) =P_{1h}(k) + P_{2h}(k) \,,
\end{equation} 
where the 1-halo and 2-halo terms are
\begin{eqnarray}
   P_{1h}(k) &=& \int dM\,{dn\over dM} \, 
	[ R_s^3\,\deltabar\,\ut (kR_s)]^2  \nonumber\\
     \noalign{\smallskip}
   P_{2h}(k) &=& \int dM\,{dn\over dM} \,R_s^3\,\deltabar\,\ut(kR_s)  
    \int dM'\,{dn\over dM'} \,R'^3_s\,\deltabar'\,\ut(kR'_s)\,P_\halo(k)
	\label{Pk} \\
    &=& \left[ \int dM\,{dn\over dM} 
	\,R_s^3\,\deltabar\,\ut(kR_s)\,b(M) \right]^2\,P_\lin(k)\,.\nonumber
\end{eqnarray}
To arrive at the last expression above, we have again used the bias
model of \eq{Pbias}.  For computational efficiency, we find that the
algebraic expressions
\begin{eqnarray}
     \ut_I(q) &=&    {4\pi \{ \ln(e+1/q) - \ln[\ln(e+1/q)]/3 \}
	\over (1 + q^{1.1})^{(2/1.1)} } \,, \qquad  p=1 \nonumber\\
     \ut_{II}(q) &=& {4\pi \{ \ln(e+1/q) + 0.25\ln[\ln(e+1/q)]\} 
 	 \over 1 + 0.8\, q^{1.5}} \,, \qquad  p={3\over 2}
\label{uq}
\end{eqnarray}
provide excellent fits for the profiles of Navarro et al. (1997) and
Moore et al. (1999), with less than 4\% rms error for form I and less
than 1\% rms error for form II.  The functional form is chosen to
reproduce the asymptotic behaviors: $\ut \sim 4\pi\ln q$ at small $q$
(with no radial cutoff), and $\ut \propto q^{-2}$ (type I) and
$\ut \propto q^{-3/2}$ (type II) at large $q$.

The two-point $\xi(r)$ and $P(k)$ can now be computed analytically
from equations~(\ref{xi2}) and (\ref{Pk}).  The inputs are \eq{u} or
(\ref{uq}) for the halo density profile $u(x)$ or $\ut(q)$,
equations~(\ref{Rs}) and (\ref{delbar}) for $R_s$ and $\bar\delta$,
\eq{PS} for the halo mass function $dn/dM$, and \eq{Pbias} for the
halo-halo correlation function.  Since the halo density profile
appears to have a nearly universal form regardless of background
cosmology, $\xi(r)$ and $P(k)$ depend on cosmological parameters
mainly through $\sigma(M)$ of \eq{sigma} and the halo concentration
$c(M)$ or central density $\deltabar(M)$.  (See Ma \& Fry 2000c for a
more detailed discussion of $c(M)$.)

\section{Three-Point Statistics: $\zeta$ and $B$}
Here we construct our analytic halo model for the three-point
correlation function $\zeta$ and the bispectrum $B$.  The joint
probability of finding three objects in volume elements $dV_1, dV_2$,
and $dV_3$ is given by
\begin{equation}
  dP=[ 1 + \xi(r_1) + \xi(r_2) + \xi(r_3) + \zeta(r_1,r_2,r_3)] 
	\,\bar{n}^3 dV_1\,dV_2\,dV_3\,,
\end{equation}
where $\xi(r)$ and $\zeta(r_1,r_2,r_3)$ are the two- and three-point
correlation functions, respectively, $\bar{n}$ is the mean number
density of objects, and $r_1, r_2$ and $r_3$ are the lengths of the
sides of the triangle defined by the three objects (Peebles 1980).
The Fourier transform of the three-point correlation function
$\zeta(r_1,r_2,r_3)$ is the bispectrum $B(k_1,k_2,k_3)$, which is
related to the density field in $k$-space by $\lexp \delta(\k_1)
\delta(\k_2) \delta(\k_3) \rexp = B(k_1,k_2,k_3)\, (2\pi)^3 \delta_D
(\k_1 +\k_2 +\k_3)\,$.  The bispectrum depends on any three parameters
that define a triangle in $k$-space.  A particular simple
configuration to study is the equilateral triangle ($k_1=k_2=k_3=k$),
and in this case the bispectrum $B^{\rm eq}$ depends only on a single
wavenumber.

Similar to the two-point halo model of \S 3, we can write the
contributions to the three-point correlation function $\zeta$ of the
mass density as three separate terms, each term representing particle
triplets that reside in a single halo, two distinct halos, or three
distinct halos.  Taking into account the halo mass distribution and
halo-halo correlations discussed in \S 2, we obtain
\begin{equation}
   \zeta(r_1,r_2,r_3) = \zeta_{1h}(r_1,r_2,r_3) + 
    \zeta_{2h}(r_1,r_2,r_3) + \zeta_{3h}(r_1,r_2,r_3)\,,
\end{equation}
where the separate 1-halo, 2-halo, and 3-halo terms are
\begin{eqnarray}
  \zeta_{1h}(r_1,r_2,r_3) &=& \int d^3 r \int dM\,{dn\over dM}\,
	\deltabar^3\, u(r/R_s)\,u(|\r+\r_1-\r_2|/R_s)\,
	u(|\r+\r_1-\r_3|/R_s) \nonumber\\
	\noalign{\smallskip}	
  \zeta_{2h}(r_1,r_2,r_3) &=& \int d^3 r\, d^3 r'
      \int dM \, {dn\over dM}\,\deltabar^2\,u(r/R_s)\,
      u(|\r+\r_1-\r_2|/R_s) 
      \int dM' \, {dn\over dM'}\,\deltabar'\, u(r'/R'_s) \nonumber \\
     && \times\, \xi_\halo(|\r-\r'+\r_1-\r_3|) +  \hbox{sym.(1,2,3)} 
	\label{zeta_123} \\   \noalign{\smallskip}
  \zeta_{3h}(r_1,r_2,r_3) &=& \int d^3 r\,d^3 r'\,d^3 r''\,
      \int dM\, {dn\over dM}\,\deltabar\, u(r/R_s) 
      \int dM'\, {dn\over dM'}\,\deltabar'\, u(r'/R'_s) \nonumber\\
    && \times \int dM''\, {dn\over dM''}\,\deltabar''\, u(r''/R_s'')
	\,\,\zeta_{\rm halo}(\r+\r_1,\r'+\r_2,\r''+\r_3)\,. \nonumber
\end{eqnarray}
The dominant contribution to the three-point correlation function in
the nonlinear regime is from the first term $\zeta_{1h}$, which comes
from particle triplets that reside in the same halo.  This term is
determined by the convolution $\gamma(\x_1,\x_2)=\int d^3 y \,u(y)\,
u(|\y + \x_1|)\, u(|\y + \x_2|)$ of three factors of the density
profile $u(x)$, and is analogous to the convolution $\lambda$ in
\eq{lamb} for the one-halo term $\xi_{1h}$ in the two-point
correlation function.

The bispectrum of the mass density field $\delta$ in $k$-space can be
obtained by Fourier transforming the equations above.  We find
\begin{equation}
   B(k_1,k_2,k_3) = B_{1h}(k_1,k_2,k_3) 
   + B_{2h}(k_1,k_2,k_3) + B_{3h}(k_1,k_2,k_3) \,,
\end{equation}
where
\begin{eqnarray}
   B_{1h}(k_1,k_2,k_3) &=& \int dM\,{dn\over dM}
    \, [R_s^3\,\deltabar\,\ut(k_1 R_s)] \, [R_s^3\,\deltabar\,\ut(k_2 R_s)]
    \, [R_s^3\,\deltabar\, \ut(k_3 R_s)] \nonumber \\
   \noalign{\smallskip}
   B_{2h}(k_1,k_2,k_3) &=& \int dM\,{dn\over dM}\, 
	[R_s^3\,\deltabar\,\ut(k_1 R_s)]\, [R_s^3\,\deltabar\,\ut(k_2 R_s)]
	\nonumber\\
       && \times \int dM'\,{dn\over dM'}\,R'^3_s\,\deltabar'\,\ut(k_3 R'_s)
	\, P_\halo(k_3; M, M') + \hbox{sym.(1,2,3)} \\
   \noalign{\smallskip}
   B_{3h}(k_1,k_2,k_3) &=& 
      \int dM \,{dn\over dM}\,R_s^3 \,\deltabar\,\ut(k_1R_s) 
      \int dM' \,{dn\over dM'}\,R'^3_s \,\deltabar'\,\ut(k_2R'_s) \nonumber \\
     && \times \int dM''\,{dn\over dM''}\,R''^3_s \,\deltabar''\,\ut(k_3R''_s) 
     \, B_\halo(k_1,k_2,k_3;M,M',M'') \,.  \nonumber
\end{eqnarray}
The halo-halo power spectrum $P_\halo(k)$ and bispectrum
$B_\halo(k_1,k_2,k_3)$ are related to the linear mass power spectrum
$P_\lin(k)$ by equations~(\ref{Pbias}) and (\ref{Bbias}).

The expressions for the mass bispectrum above simplify considerably
for the equilateral triangle configuration, and
\begin{equation}
   B^{\rm eq}(k) = B^{\rm eq}_{1h}(k) + B^{\rm eq}_{2h}(k) 
    + B^{\rm eq}_{3h}(k) \,,
\end{equation}
where
\begin{eqnarray}
   B^{\rm eq}_{1h}(k) &=& \int dM\,{dn\over dM}\,
       [R_s^3\,\deltabar\,\ut (kR_s)]^3 \nonumber\\
   B^{\rm eq}_{2h}(k) &=& 3\, \left[ \int dM\,{dn\over dM}\,
	[R_s^3\,\deltabar\,\ut(kR_s)]^2\,b(M)\right]
     \left[\int dM\,{dn\over dM}\,R_s^3\,\deltabar\,\ut(kR_s)\,b(M) \right]
        \, P_{\rm lin}(k)\,   \label{Bk} \\
   B^{\rm eq}_{3h}(k) &=& \left[ \int dM\,{dn\over dM}\,
        R_s^3\,\deltabar\,\ut(kR_s)\,b(M) \right]^3\,
	{12\over 7}\,P^2_{\rm lin}(k)\,.\nonumber
\end{eqnarray}
Here we have written out explicitly the bias factors $b(M)$ using
equations~(\ref{Pbias}) and (\ref{Bbias2}), and we have neglected
terms with $b_2(M)$ as discussed in \S 2.3.

\section{$N$-body Experiments and Numerical Results}

In this section we compare the predictions of our analytical model
described in \S 2, 3, and 4 with results from cosmological $N$-body
simulations.  We examine two cosmological models: an $n=-2$ scale-free
model and a low-density $\Lambda$CDM model.  These are the same
simulations studied in Ma \& Fry (2000a).  The $n=-2$ simulation has
$256^3$ particles and a Plummer force softening length of $L/5120$,
where $L$ is the box length.  The $\Lambda$CDM model is spatially flat
with matter density $\Omega_m=0.3$ and cosmological constant
$\ov=0.7$.  This run has $128^3$ particles and is performed in a $
(100\,{\rm Mpc})^3$ comoving box with a comoving force softening
length of $ 50\,{\rm kpc} $ for Hubble parameter $ h=0.75 $.  The
baryon fraction is set to zero for simplicity.  The primordial power
spectrum has a spectral index of $n=1$, and the density fluctuations
are drawn from a random Gaussian distribution.  The gravitational
forces are computed with a particle-particle particle-mesh (P$^3$M)
code (Ferrell \& Bertschinger 1994).  We compute the density field
$\delta$ on a grid from particle positions using the second-order
triangular-shaped cloud (TSC) interpolation scheme.  A fast Fourier
transform is then used to obtain $\delta$ in $k$-space.  The $k$-space
TSC window function is deconvolved to correct for smearing in real
space due to the interpolation, and shot noise terms are subtracted to
correct for discreteness effects.  We then compute the second and
third moments of the density amplitudes in Fourier space.

We show results for the power spectrum as the dimensionless variance $
\Delta(k) \equiv 4 \pi k^3 P(k)/(2\pi)^3 $.  A useful dimensionless
three-point statistic is the hierarchical three-point amplitude
\begin{equation} 
	Q(k_1,k_2,k_3) \equiv { B(k_1,k_2,k_3)\over P(k_1) P(k_2) 
	+ P(k_2) P(k_3) + P(k_3) P(k_1)} \,.
\label{Qdef}
\end{equation}
The three-point amplitude $Q$ has the convenient feature that for the
lowest nonvanishing result in perturbation theory, $Q$ is independent
of time and the overall amplitude of $P$; for scale-free models with a
power-law $P$, $Q$ is independent of overall scale as well.  To
lowest order, it follows from \eq{Btree} that the equilateral
bispectrum has a particularly simple form, $B^{(0)}(k) = {12\over 7}\,
P_\lin^2(k)\,$, and we have $Q^{(0)}(k)={4\over 7}\,$, independent of
the power spectrum.

\subsection{Synthetic Halo Replacement}

To investigate the numerical effects of limited resolution in the
simulations, we have experimented with the distribution of matter in
halos identified in the simulations.  In these experiments, we keep
the locations and masses of the halos unchanged but redistribute the
subset of particles which lies within the virial radius $R_{200}$ (the
radius within which the mean overdensity is 200) of each halo
according to a prescribed density profile.  We then recompute the two-
and three-point statistics $\Delta$ and $Q$ from the redistributed
particle positions as well as the original non-halo particles, which
remain at their original positions.  By using density profiles
obtained empirically from higher-resolution simulations of individual
halos, this recipe allows us to model accurately the inner regions of
the halos on scales below the numerical softening length scale while
at the same time preserving all the large-scale information available
in the large parent simulation.  This technique should also be useful
for other studies that are sensitive to the inner halo density
profiles, for example the ray-tracing method in gravitational lensing.

Ma \& Fry (2000a) have used this replacement technique to experiment
with synthetic halos that follow a pure power-law profile $u \propto
r^{-\epsilon}$.  It is found that $\Delta(k)$ and $Q(k)$ at high-$k$
indeed obey $\Delta(k) \propto k^{2\epsilon-3}$ and $Q(k)\propto
k^{3-\epsilon}$ as predicted by the simple power-law model of Peebles
(1974).  The scaling works even in the presence of the full
distribution of matter outside the halo cores.  Here we extend this
replacement technique to more realistic halo profiles of
equation~(\ref{u}).  Figures 1 and 2 illustrate the effects on the
matter power spectrum and bispectrum when the original halos in large
cosmological simulations are replaced by synthetic halos with the
density profile $u_{II}=1/(x^{3/2}+x^3)$ of equation~(\ref{u}).  For
the $n=-2$ scale-free model, the concentration parameter is taken to
be $c(M)=3(M_*/M)^{1/6}$, which is consistent with Navarro et
al. (1997) and has the expected scaling with mass, $c\propto
M^{-(3+n)/6}$, in a scale-free model.  For $\Lambda$CDM models, we use
$c(M)=5(M_*/M)^{1/6}$ as suggested by Figure~3 of Moore et al. (1999).
We note, however, that $c(M)$ from various recent simulations has
shown a large scatter, and its functional form depends on the exact
form of the density profile used.  For the $\Lambda$CDM model and form
$u_{II}=1/(x^{3/2}+x^3)$, for example, a flatter and smaller
$c(M)=3(M_*/M)^{0.084}$ appears to be preferred by Jing \& Suto (2000)
and Navarro et al. (1997).  The results of Tormen et al. (1997) and
Cole \& Lacey (1996) are also only marginally consistent with each
other.  A more detailed investigation of the different forms of $c(M)$ 
can be found in Ma \& Fry (2000c).

In Figures 1 and 2, the agreement at low values of $k$ between the
original and synthetic halos is excellent, confirming that the
correlation functions on larger length scales are insensitive to the
spatial distribution of particles in the halo cores.  The only
significant difference between the simulation and synthetic halos
appears at small length scales, where the coarser resolution of the
simulation blurs out the structure of the inner halo and results in an
inner profile flatter than in equation~(\ref{u}).  This effect is
manifested in the bending over of the dashed curves for $P(k)$ in
Figures 1 and 2 at high $k$, and is corrected for when the synthetic
halos are used.

\subsection{$N$-body Results vs. Analytic Halo Model}

We now proceed to compare the predictions of the analytic model of \S
2 -- \S 4 with the numerical results from cosmological simulations.
Figures~3 and 4 show the $k$-space density variance $\Delta(k) $
(upper panel) and the three-point amplitude $Q_{\rm eq} (k)$ for
equilateral triangles for the $n=-2$ scale-free model and the
$\Lambda$CDM model.  The solid black curves are the model predictions
computed from equations~(\ref{Pk}) and (\ref{Bk}).  The contribution
from the single-halo and multiple-halo terms are shown separately as
dashed curves.  For the density profile, we use the same
$u_{II}=1/(x^{3/2}+x^3)$ and concentration parameters as in Figures~1
and 2.  For the mass function, we use the Press-Schechter formula but
reduce its overall amplitude by 25\%, which we find necessary in order
to match the halo mass functions for our numerical simulations.  This
overestimation of halo numbers with $M\sim M_*$ by Press-Schechter is
a well known result reported in many other studies (see Jenkins et
al. 2000 and references therein).  The mass limits for the integrals
in equations~(\ref{Pk}) and (\ref{Bk}) do not significantly affect the
model predictions for the total $\Delta$ or $Q$.  Raising the lower
mass limit does reduce the contribution from lower mass halos and
hence lower the high-$k$ amplitudes of the multiple halo terms
$\Delta_{2h}$, $Q_{2h}$, and $Q_{3h}$, but these terms make negligible
contributions to the total $\Delta$ and $Q$.

As discussed in \S 3 and 4, the nonlinear parts of both the two- and
three-point statistics are determined by the dominant 1-halo term
because the closely spaced particle pairs and triplets mostly reside
in the same halos.  The multiple-halo terms are therefore significant
only on larger length scales comparable to the separation between
halos.  Their inclusion, however, is necessary for the transition into
the linear regime.

For the $n=-2$ model in Figure~3, we plot the results against the
scaled $k/k_\nl$, where $k_\nl$ characterizes the length scale that is
becoming nonlinear and is defined by $\int_0^{k_\nl} d^3k\,
P_\lin(a,k)/(2\pi)^3=1$.  Three time outputs are shown, where the
expansion factor (1 initially) and $k_\nl$ (in units of $2\pi/L$) are:
$(a,k_\nl)= (13.45, 29), (19.03, 14.5)$, and $(26.91, 7.25)$ (from
left to right).
For the two-point $\Delta(k)$, the agreement between the model
prediction and the simulations is excellent.  The three simulation
outputs also overlap well, indicating that self-similarity is obeyed,
as reported in Jain \& Bertschinger (1998).  For the three-point
$Q_{\rm eq}$, however, self-similar scaling does not hold as
rigorously (Ma \& Fry 2000a).  It is interesting to note that the
analytic prediction agrees most closely with the earliest output
$(a,k_\nl)= (13.45, 29)$ (green curve).  This provides further
evidence to the suggestion of Ma \& Fry (2000a) that the later outputs
of the $n=-2$ simulation may be affected by the finite volume of the
simulation box.  For the $\Lambda$CDM model in Figure~4, the analytic
model again provides a good match to the $N$-body results within the
fluctuations among the simulations.  We illustrate the numerical
effects due to box sizes by showing results from two runs with volume
(100 Mpc)$^3$ and (640 Mpc)$^3$.  The model predictions extend well
beyond the resolution of the simulations.  

The real-space two-point correlation function for the $n=-2$ and
$\Lambda$CDM models is shown in Figures~5 and 6.  For the halo model
predictions, we have chosen to show only the results for the 1-halo
term $\xi_{1h}$ because this term dominates the interesting nonlinear
portion of $\xi$.  The agreement between the halo model (dashed
curves) and the simulations (symbols) is again excellent.  For the
2-halo terms $\xi_{2h}$, the computation can be done more easily in
$k$-space as shown in Figures~3 and 4, so we do not include them here.

For comparison, we plot in Figures~3--6 the results from the commonly
used fitting formulas for the nonlinear power spectrum (Hamilton et
al. 1991; Jain et al. 1995; Peacock \& Dodds 1996; Ma 1998; Ma et
al. 1999).  While the formulas provide a good approximation to
$\Delta(k)$ up to $k/k_\nl \sim 50$ for the $n=-2$ model and $k\sim
20\,h$ Mpc$^{-1}$ for the $\Lambda$CDM model, the figures show that
significant deviations occur at higher $k$, and the fitting formula
and our current model predict different high-$k$ slopes for
$\Delta(k)$.  Since the high-$k$ behavior of the fitting formulas has
been constructed to obey the stable clustering prediction, this
discrepancy has an important implication for the validity of stable
clustering, which we discuss briefly in the next section and at length
in Ma \& Fry (2000c).

\section{Discussion}

We have constructed a physical model for the correlation functions of
the mass density field in which the correlations are derived from
properties of dark matter halos.  We have described in detail the
input, construction, and results of this model in \S 2 -- \S 5.  We
now examine more closely its physical meanings and implications in
three separate regimes.  

On scales larger than the size of the largest halo, the contributions
from separate halos dominate, and (by design) the model reproduces the
results of perturbation theory.  On intermediate scales, $ 1/R_* \la k
\la 1/R_s(M_*) $, because of the exponential cutoff in the mass
function $dn/dM$ at the high mass end, the contribution to the volume
integrals in \eq{xi2} is dominated by the large-$r$ regime where the
halo profiles are roughly $r^{-3}$.  The correlation functions
therefore behave approximately as predicted by the power-law model
with $\epsilon=3$, i.e., $\Delta \propto k^{2\epsilon-3} \sim k^3$ and
$Q \propto k^{3-\epsilon} \sim $ constant.  This is why $Q$ exhibits
an approximately flat plateau at intermediate $k$ in the bottom panels
of Figures~3 and 4.

On the smallest and most nonlinear scales, the correlation functions
probe the innermost regions of the halos.  Intriguingly, the halo
model predicts on these scales a behavior that is different from
either the frequently-assumed stable clustering result of $\Delta(k)
\propto k^\gamma $ with $ \gamma = (9+3n)/(5+n)$ (Davis \& Peebles
1977), or the power-law profile result of $\gamma = 2\epsilon - 3$.
The implication of departure from stable clustering is significant
because all the fitting formulas for the nonlinear $P(k)$ in the
literature (see \S 5.2) have been constructed to approach the stable
clustering limit at high $k$.  A more detailed study on the criteria
for stable clustering in this model is given in a separate paper (Ma \&
Fry 2000c).

The origin of the deviation from stable clustering in the model at
high-$k$ can be understood as follows.  For the two-point function, as
$k$ becomes large, the one-halo integral $P_{1h}(k)$ in \eq{Pk}
converges before the exponential cutoff, and is dominated by
contributions near the mass scale for which $kR_s=1 $.  The behavior
now depends on the mass distribution function.  The various mass
functions discussed in \S 2.2 have the same general behavior of $dn/dM
\propto M^{-2}\,\nu^\alpha\, e^{-\nu^2/2}$, where
$\nu=\delta_c/\sigma$.  The Press-Schechter form assumes $\alpha=1$
(see eq.~[\ref{PS}]), while others (e.g., Sheth \& Tormen 1999;
Jenkins et al. 2000) suggest a flatter slope of $\alpha\approx 0.4$
for the lower mass halos.  Since the scale radius $R_s$ depends on
mass as $R_s = R_{200}/c \propto M^{1/3} / M^{-(3+n)/6} \propto
M^{(5+n)/6}$, and $R_s^3 \deltabar \propto M $ (up to logarithmic
factors), we find from \eq{Pk} that the power spectrum at high $k$
goes as
\begin{equation}
  \Delta(k) \approx \Delta_{1h}(k) \propto k^3\,\int dM\,\nu^\alpha
  \, \ut^2(kR_s) \,.
\end{equation}
Changing variables to $y = kR_s \propto k\,(M/M_*)^{(5+n)/6}$, we
see that
\begin{equation}
   \Delta(k)\propto k^\gamma\,,\qquad 
	\gamma = \left( {9+3n\over 5+n} \right) - 
        \alpha \left( {3+n\over 5+n} \right) \,, \label{g2halo} 
\end{equation}
where the first term in $\gamma$ is the prediction of stable
clustering.  The departure arises from the factor $\nu^\alpha$ in the
mass function, and would vanish only if $\alpha=0$ or $n=-3$.  This is
the origin of the difference in $\Delta(k)$ at high $k$ between the
model prediction (solid curves) and the fitting formula (dotted
curves) shown in Figures 3 and 4.

For the three-point function, the one-halo integral $B^{\rm eq}_{1h}$
in \eq{Bk} converges (barely, for $ p = {3 \over 2} $ and $ n =-2 $),
giving
\begin{eqnarray}
     B^{\rm eq}(k) &\propto & k^{\gamma_3 -6}\,,\qquad 
    \gamma_3 = 2 \left( {9+3n \over 5+n} \right) - 
    \alpha \left( {3+n\over 5+n} \right) \nonumber\\
     Q^{\rm eq}(k) &\propto & k^{\alpha (3+n)/(5+n)}
\label{g3halo} 
\end{eqnarray}
This again disagrees with the prediction of stable clustering that $Q$
is constant, but it appears to be consistent with numerical
simulations as shown in Figures~3 and 4.  

For yet higher order correlations, details of the halo profile begin
to matter.  For $p=1$, the pattern of equations (\ref{g2halo}) and
(\ref{g3halo}) persists to all orders, but for $ p = {3 \over 2} $
they apply only for the two- and three-point functions; for four-point
and higher functions the nonlinear scale $M_* $ and $ \gamma_n = np -
3 $ for $ n \ge 4 $.  Thus there seems to be some potentially
interesting behavior that is tested only in the four-point function
and higher.

\section{Summary}

We have presented an analytic model for the two- and three-point
correlation functions $\xi(r)$ and $\zeta(r_1,r_2,r_3)$ of the
cosmological mass density field and their Fourier transforms, the mass
power spectrum $P(k)$ and the bispectrum $B(k_1,k_2,k_3)$.  In this
model, the clustering statistics of the density field are derived from
a superposition of dark matter halos with a given set of input halo
properties.  These input ingredients include realistic halo density
profiles of \eq{u}, halo mass distribution of \eq{PS}, and halo-halo
spatial correlations of equations~(\ref{Pbias}) and (\ref{Bbias2}).
The main results of the model are given by equations~(\ref{xi2}) and
(\ref{Pk}) for the two-point statistics $\xi$ and $P$, and by
equations~(\ref{zeta_123}) and (\ref{Bk}) for the three-point
statistics $\zeta$ and $B$.  This model provides a rapid way to
compute the correlation functions over all length scales where the
model inputs are valid; it also gives a physical interpretation of the
clustering process of matter in the universe.

We have tested the validity of this model by comparing its predictions
with results from cosmological simulations of an $n=-2$ scale-free
model and a $\Lambda$CDM model.  As Figures~3 -- 6 illustrate, the
model describes well the simulation results spanning the entire range
of behavior from the perturbative regime on large scales to the
strongly nonlinear regime on small scales.  To probe the critical
high-$k$ range in the deeply nonlinear regime, we have used a halo
replacement technique to increase the resolution of the large parent
simulations.  As Figures~1 and 2 illustrate, this method of replacing
the original halos that suffer from numerically softened cores with
synthetic halos of analytic profiles is a reasonable way to improve
the resolution of numerical simulations.  By using density profiles
obtained empirically from higher-resolution simulations of individual
halos, this recipe allows us to model accurately the inner regions of
the halos on scales below the numerical softening length scale, while
at the same time preserving all the large-scale information available
in the large parent simulation.  This technique should also be useful
for other studies that depend on the inner halo density profiles, for
example, the ray-tracing method in gravitational lensing.

Given that dark matter halos in simulations (and presumably in nature)
are not perfectly spherical, cleanly delineated objects, it is
intriguing that the model constructed in this paper works as well as
it does at matching the simulation results.  Nevertheless, this
analytic model provides a good qualitative and quantitative
description over the entire range of scales covered by the simulation,
and it can be used to make predictions beyond these scales.  This is
the first model prescription that successfully reproduces both two-
and three-point mass correlations.  We believe that it will prove to
be a generally useful framework.

\acknowledgements

We have enjoyed stimulating discussions with John Peacock and David
Weinberg.  We thank Edmund Bertschinger for valuable comments and for
providing the $n=-2$ scale-free simulation.  Computing time for this
work is provided by the National Scalable Cluster Project and the
Intel Eniac2000 Project at the University of Pennsylvania.
C.-P. M. acknowledges support of an Alfred P. Sloan Foundation
Fellowship, a Cottrell Scholars Award from the Research Corporation, a
Penn Research Foundation Award, and NSF grant AST 9973461.

\clearpage

\appendix
\section{Appendix}

In this Appendix we display analytic forms for the convolution of
the dimensionless profile shape
\begin{equation} 
   \lambda(x) = \int d^3y \, u(y)\, u(|\x +\y|) 
\label{lamb2}
\end{equation}
discussed in \S 3.  These analytic expressions are useful for
computing the nonlinear two-point correlation function $\xi$ of the
mass density field, which is dominated by the 1-halo term $\xi_{1h}$
in \eq{xi2} and is related to $\lambda$ by
\begin{equation} 
   \xi(r)\approx \xi_{1h}(r) = \int dM {dn\over dM}\,\deltabar^2\,
	R_s^3\,\lambda(r/R_s) \,,\qquad {\rm for\ } \xi \ga 1 \,.
\end{equation}

For the type-I profile $u_I$ of \eq{u}, the angular integration in
equation~(\ref{lamb2}) is analytic, and $\lambda$ is reduced to a
simple integral
\begin{equation}
   \lambda_I(x)={2\pi \over (2-p)x} \int_0^\infty 
   {y \, dy \over y^{p}\,(1+y)^{3-p}} \left[ {(x+y)^{2-p}\over (1+x+y)^{2-p}}
	-{|x-y|^{2-p} \over (1+|x-y|)^{2-p} } \right] \,.
\label{lamb1a}
\end{equation}
For the special case $p=1$, this integral can
be further reduced to the analytical form
\begin{equation}
   \quad \lambda_I(x)={8\pi \over x^2(x+2)} \left[ {(x^2+2x+2)\,\ln(1+x)
	\over x\,(x+2)} - 1 \right]\,, \qquad p=1 \,.
\label{lamb1b}
\end{equation}
For $u_{II}$ of \eq{u}, we are able to simplify $\lambda$ to
\begin{equation}
   \lambda_{II}(x)={2\pi \over x} \int_0^\infty 
   { y\, dy \over y^p(1 + y^{3-p})} \, F_p(x,y) \,,
\end{equation}
where the function $F_p(x,y)$ represents the angular part 
of the integration in equation~(\ref{lamb2}) and
\begin{equation}
    F_p(x,y)=\int_{|x-y|}^{x+y} {z \, dz \over z^{p}(1+z^{3-p})}\,. 
\label{Fp}
\end{equation}
The integral in $F_p$ can be reduced to analytic forms for special
values of $p$.  Here we display the six cases $p=0$, $1/2$, 1, $3/2$, 2,
and $5/2$:
\begin{eqnarray}
\noalign{\smallskip}
  F_0 &=& {1 \over 6} \left\{ 2\sqrt{3} \tan^{-1} \left[{-1+ 2 (x+y)
  \over \sqrt{3}} \right] + \ln \left[ 1 - (x+y) + (x+y)^2 
  \over 1 + 2(x+y) + (x+y)^2 \right] \right\} \nonumber \\
 && \qquad -{1\over 6} \left\{ 
	\hbox{replace $(x+y)$ above with $|x-y|$} \right\} \\
	\noalign{\smallskip}
  F_{1/2} &=& {1 \over 10} \left\{ -2 \sqrt{10 + 2\sqrt5} \,
   \tan^{-1} \! \left( {1 + \sqrt5 - 4 \sqrt{x+y} \over
   \sqrt{10 - 2\sqrt5)}}\right) \right. \nonumber \\
   && \qquad -2 \sqrt{10 - 2 \sqrt5} \, \tan^{-1} \! \left( {-1 +
   \sqrt5 + 4 \sqrt{x+y} \over \sqrt{10 + 2\sqrt5)}}\right) \nonumber \\
   && + 4 \ln\left( 1 + \sqrt{x+y} \right)
   - (1 + \sqrt5) \ln\left[ 1 + {1\over 2}
   (-1 + \sqrt5) \sqrt{x+y} + x+y \right] \nonumber \\
   && \left. - (1 - \sqrt5) \ln \left[ 1 - {1\over 2} (1 + \sqrt5)
   \sqrt{x+y} + x+y \right] \right\} \nonumber \\
   && \qquad -{1\over 10} \left\{
   \hbox{replace $(x+y)$ above with $|x-y|$} \right\} \\
	\noalign{\smallskip}
  F_1 &=& \tan^{-1}(x+y)  - \tan^{-1}(|x-y|) \\
	\noalign{\smallskip}
  F_{3/2}
  &=& {1\over 3} \left\{ 2\sqrt{3} \tan^{-1} \left[{-1+ 2\sqrt{x+y}
  \over \sqrt{3}} \right] + \ln \left[ 1+2 \sqrt{x+y} + x+y \over  
  1 - \sqrt{x+y} + x+y \right] \right\}  \nonumber\\
  && \qquad -{1\over 3} \left\{ 
	\hbox{replace $(x+y)$ above with $|x-y|$} \right\} \\
	\noalign{\smallskip}
  F_2 &=& \ln \left[ {x+y \over 1+x+y} \right] - 
	\ln \left[ {|x-y| \over 1+|x-y|} \right] \\
	\noalign{\smallskip}
  F_{5/2} &=& {2 \over \sqrt{|x-y|}} - {2 \over \sqrt{x+y}}
	+ \ln\left[{ (1 + 2\sqrt{x+y} + x+y) \, |x-y| \over 
	 (1 + 2\sqrt{|x-y|} + |x-y|) \, (x+y)}\right] 
\end{eqnarray}

\clearpage

\epsscale{1.}  
\plotone{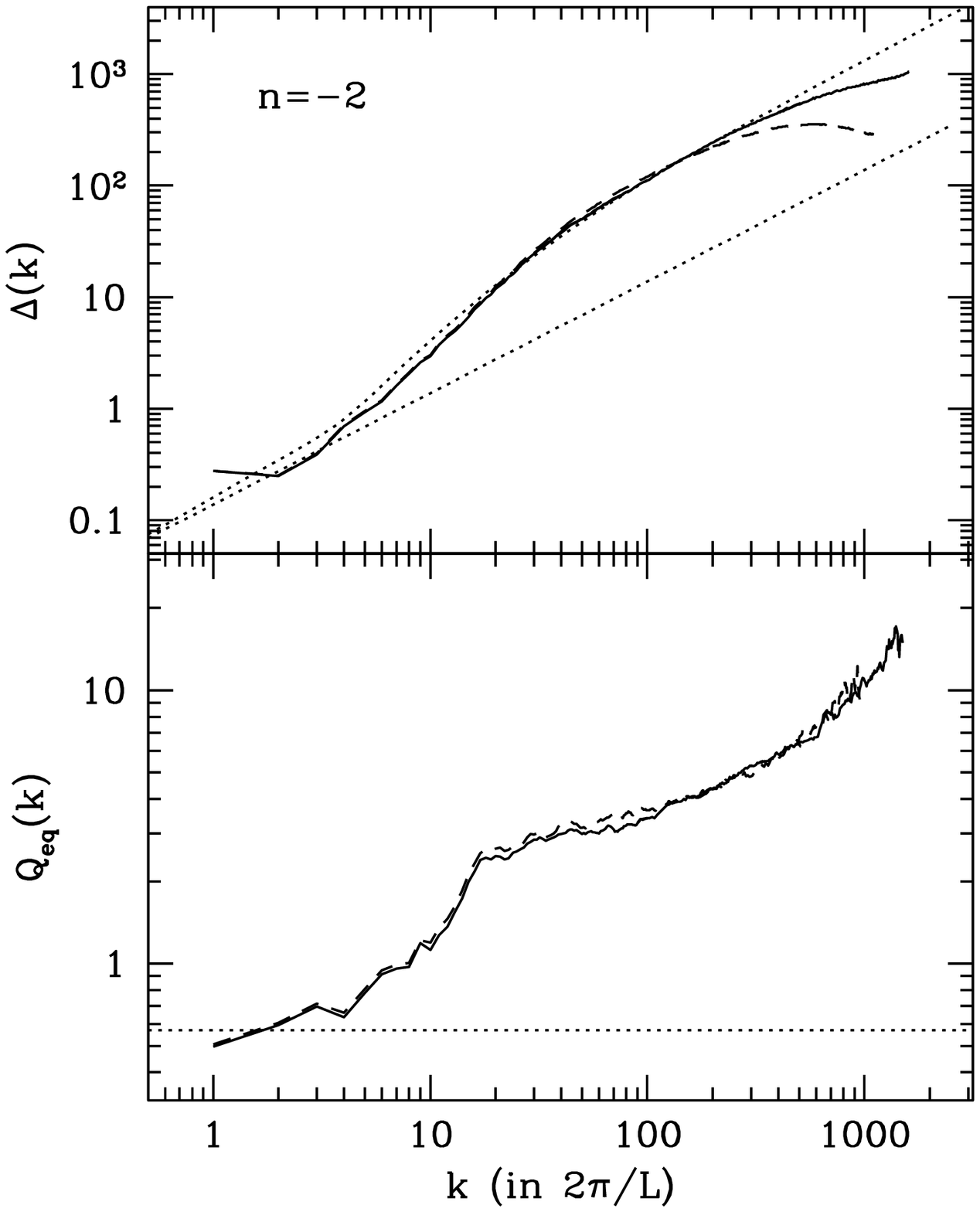}
\figcaption {Effects on the power spectrum (upper panel) and
bispectrum (lower panel) when the dark matter halos in an $n=-2$
scale-free simulation are replaced with synthetic halos of density
profile $u_{II}(x)= 1/(x^{3/2}+x^3)$ and concentration parameter
$c(M)=3\,(M_*/M)^{1/6}$ (see \S 2 for definitions).  The dashed and
solid curves are for the original and the redistributed particles,
respectively.  They agree up to $k\approx 200$ (in $2\pi/L$), beyond
which the dashed curves deviate due to the finite resolution in the
original simulation.  The dotted curves show the linear $\Delta$
and the nonlinear fitting of Jain et al. (1995) in the upper panel,
and the lowest-order perturbative result $Q^{(0)}=4/7$ in the
bottom panel.}

\plotone{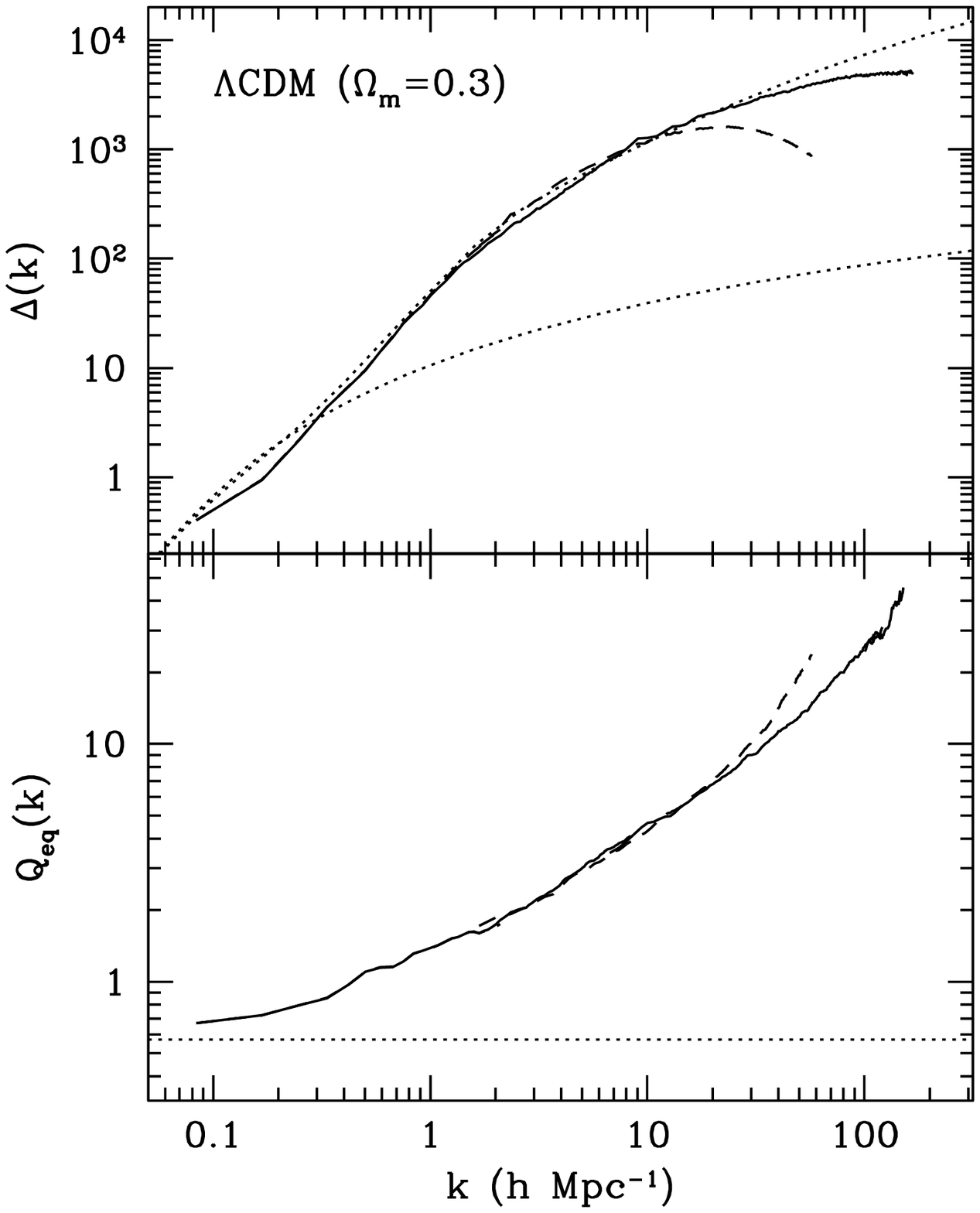}
\figcaption {Same as Fig.~1 but for a low-density CDM simulation with
$\om=0.3,\ov=0.7$.  The synthetic halos have the
$u_{II}(x)=1/(x^{3/2}+x^3)$ profile and concentration parameter
$c(M)=5\,(M^*/M)^{1/6}$.  Again, the original (dashed) and
redistributed (solid) particles have similar $\Delta(k)$ and $Q_{\rm
eq}(k)$ up to the simulation resolution of $k\approx
20\,h\,$Mpc$^{-1}$. }

\epsscale{.95}  
\plotone{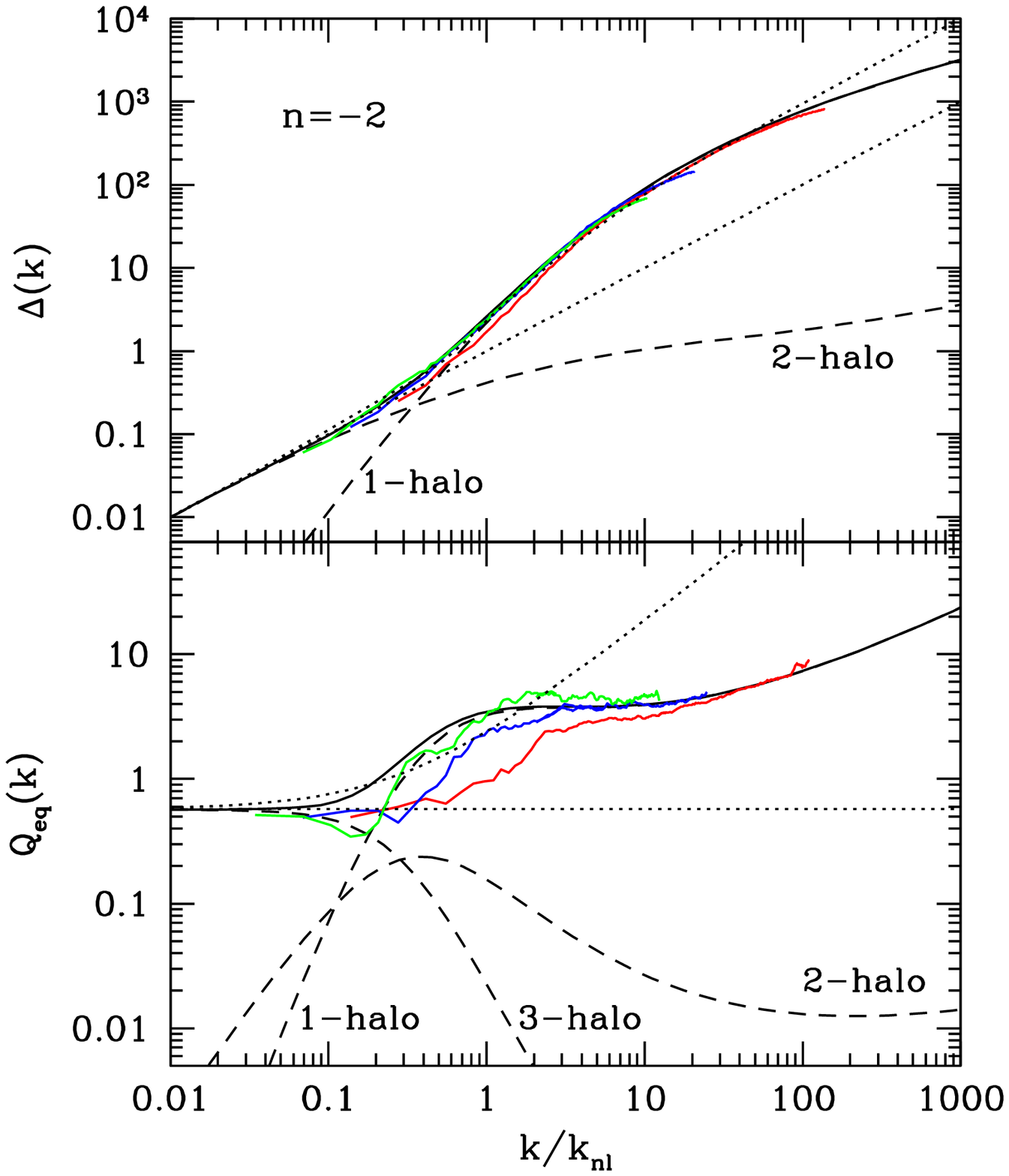} 
\figcaption {$N$-body results vs. predictions of the analytic model of
\S 2--\S 4 for the power spectrum (upper) and bispectrum (lower) for
the $n=-2$ scale-free model.  The dashed curves show the separate
contributions to $\Delta$ and $Q_{\rm eq}$ computed from the single-
and multiple-halo terms of eqs.~(\protect{\ref{Pk}}) and
(\protect{\ref{Bk}}); the solid black curves show the sum predicted by
the model.  The colored curves show the $N$-body results, where
synthetic halos have been used to extend the curves to higher $k$.
(The same density profile and $c(M)$ are used for the synthetic halos
and the analytic model.)  Three simulation outputs are shown, where
the expansion factor (1 initially) and nonlinear wavenumber (in units
of $2\pi/L$) are: $(a,k_\nl)=$ (13.45, 29), (19.03, 14.5), and (26.91,
7.25) (from left to right in green, blue and red).  Three of the four
dotted curves are the same as in Fig.~1; the rising one in the bottom
panel shows the 1-loop $Q$.}

\epsscale{1.}  
\plotone{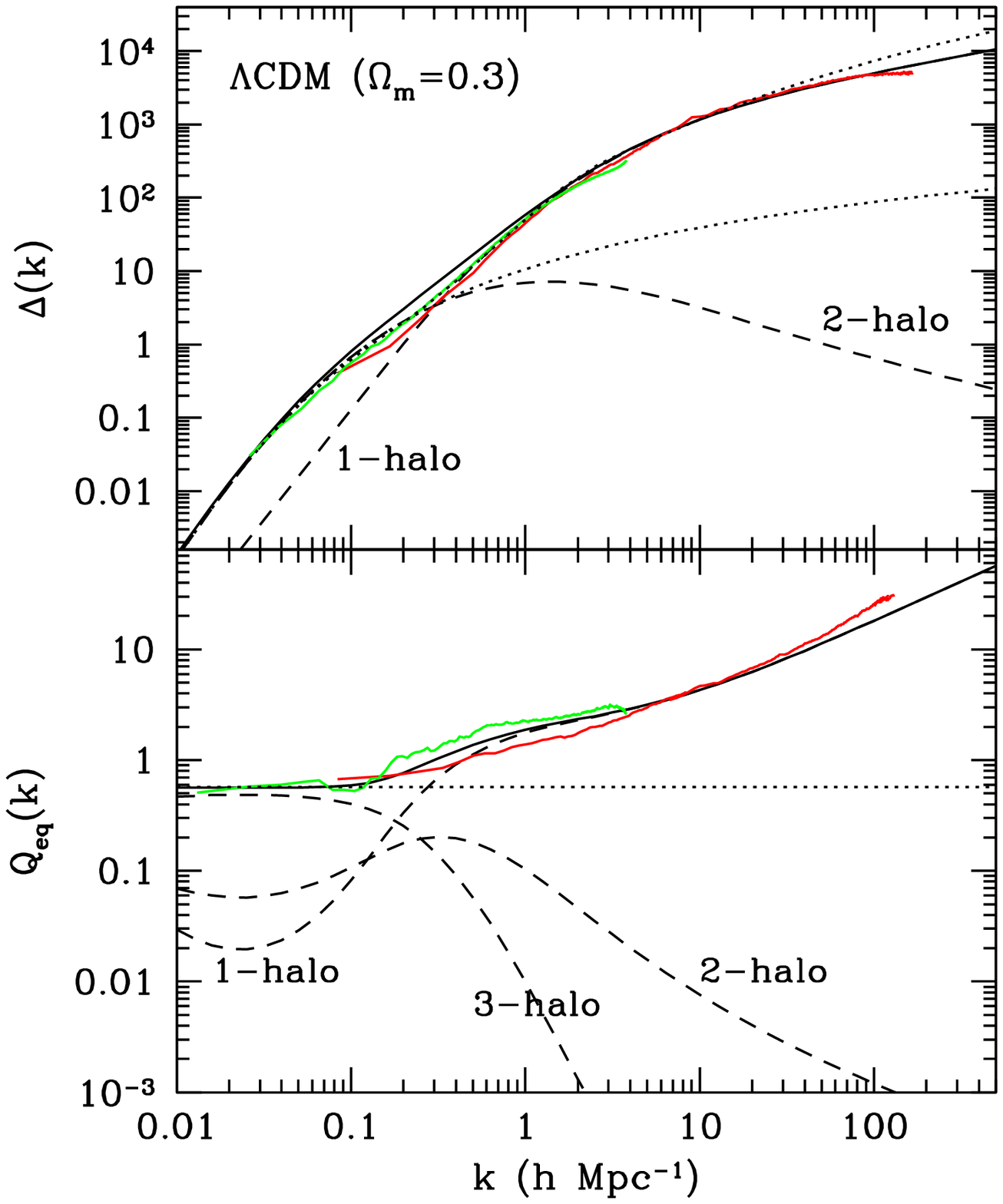} 
\figcaption {Same as Fig.~3 but for the low-density
CDM simulation with $\om=0.3,\ov=0.7$.  The red and green curves
are computed from a (100 Mpc)$^3$ and a (640 Mpc)$^3$ simulation,
respectively.}

\epsscale{1.1}  
\plotone{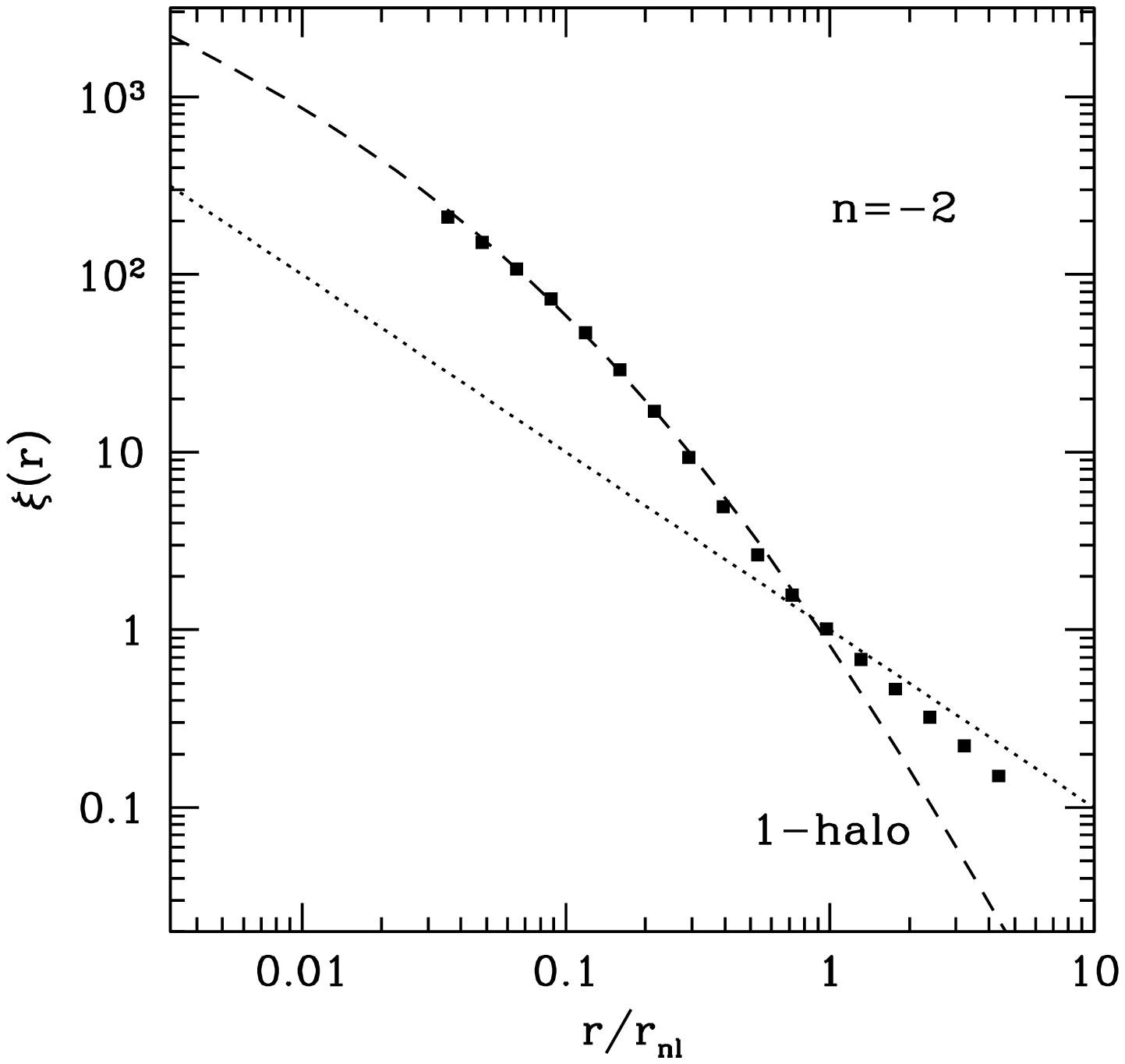} 
\figcaption {$N$-body results vs. predictions of the analytic model of
\S 2--4 for the two-point correlation function $\xi(r)$ for the $n=-2$
model.  The dashed curve shows the 1-halo term $\xi_{1h}(r)$ of
\eq{xi2} from our analytic model.  The solid squares show $\xi(r)$
computed directly from an $N$-body simulation.  The two agree very
well for $r/r_{\rm nl} \la 1$.  The dotted curve shows the linear
theory $\xi_\lin(r)=r_\nl/r$. }

\epsscale{1.1}  
\plotone{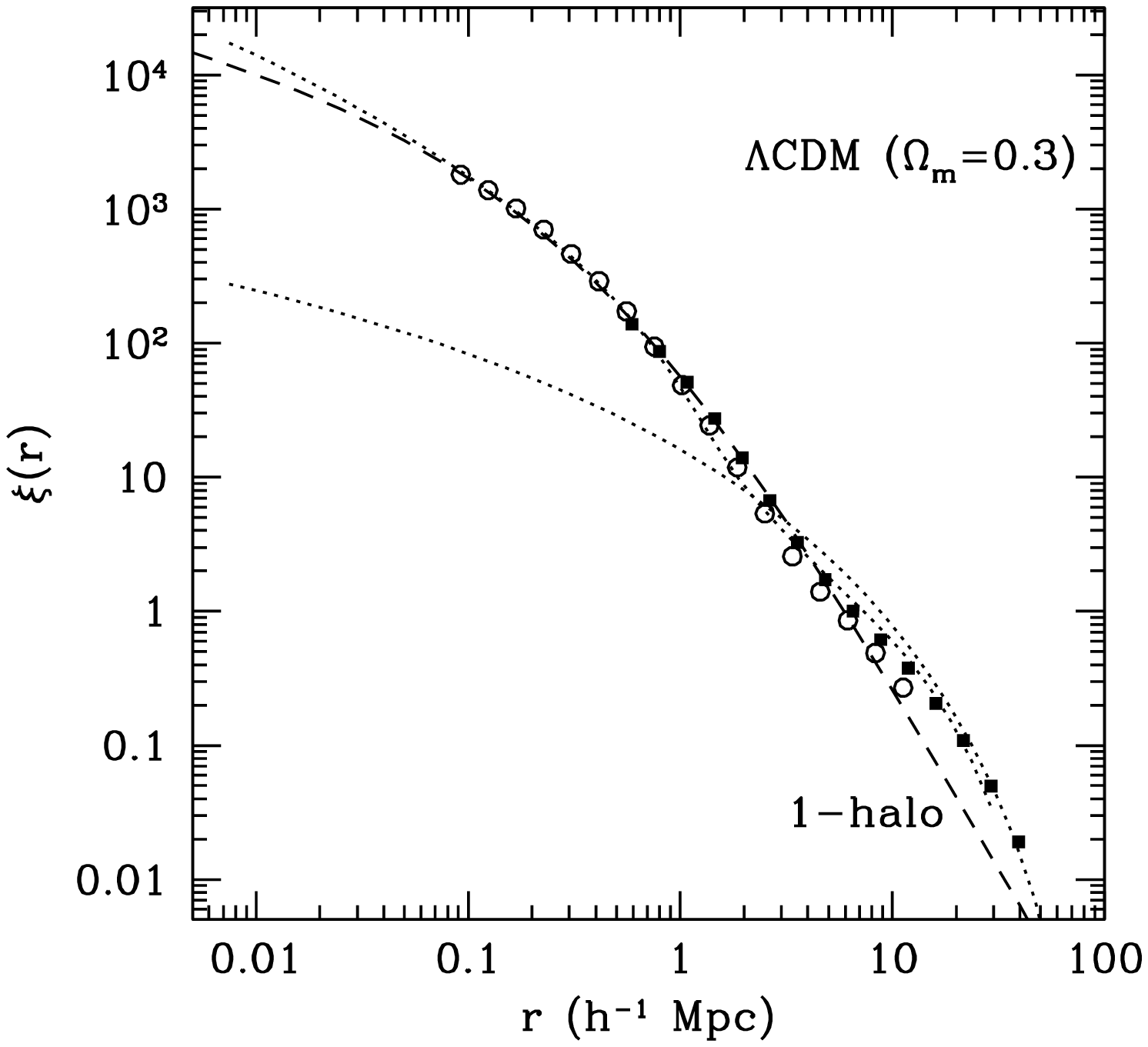} 
\figcaption {Same as Fig.~5 but for the $\Lambda$CDM model.  The
symbols show $\xi(r)$ computed from a (100 Mpc)$^3$ (open circles) and
a (640 Mpc)$^3$ (solid squares) $N$-body simulation.  The dotted
curves show $\xi_\lin(r)$ from the linear theory (lower curve) and the
nonlinear $\xi(r)$ (upper curve) given by the fitting formula of Ma
(1998).}


\clearpage

\begin{thebibliography}{}

\bibitem[Cole \& Lacey(1996)]{CL96}
Cole, S., \& Lacey, C. 1996, \mnras, 281, 716

\bibitem[Davis \& Peebles(1977)]{DP77}
Davis, M., \& Peebles, P. J. E. 1977, \apjs, 34, 425

\bibitem[Dubinski \& Carlberg(1991)]{DC91}
Dubinski, J., \& Carlberg, R.  1991, \apj, 378, 496

\bibitem[Ferrell \& Bertschinger(1994)]{FB94}
Ferrell, R., \& Bertschinger, E. 1994, Int.~J.~Mod.~Phys.~C, 5, 933

\bibitem[Fry(1984)]{F84}
Fry, J. N., 1984, \apj, 279, 499

\bibitem[Fry \& Peebles(1978)]{FP78}
Fry, J. N., \& Peebles, P. J. E. 1978, \apj, 221, 19

\bibitem[Fukushige \& Makino(1997)]{FM97}
Fukushige, T., \& Makino, J.  1997, \apj, 477, L9

\bibitem[Groth \& Peebles(1977)]{GP77}
Groth, E. J., \&  Peebles, P. J. E. 1977, \apj, 217, 385

\bibitem[Hamilton et al.(1991)]{HMKL91}
Hamilton, A. J. S., Matthews, A., Kumar, P., \& Lu, E.  1991, ApJ, 374, L1
                 
\bibitem[Hernquist(1990)]{H90}
Hernquist, L. 1990, \apj, 356, 359

\bibitem[Huss, Jain, \& Steinmetz(1999)]{HJS99}
Huss, A., Jain, B., \& Steinmetz, M. 1999, \apj, 517, 64

\bibitem[Jain \& Bertschinger(1998)]{JB98} 
Jain, B., \& Bertschinger, E.  1998, \apj, 509, 517

\bibitem[Jain, Mo, \& White(1995)]{JMW95}
Jain, B., Mo, H. J., \& White, S. D. M. 1995, \mnras, 276, L25

\bibitem[Jenkins(2000)]{J00}
Jenkins, A., Frenk, C. S., White, S. D. M., Colberg, J. M., Cole, S.,
Evrard, A. E., \& Yoshida, N.  2000, astro-ph/0005260

\bibitem[Jing(1998)]{J98}
Jing, Y. P.  1998, \apj, 503, L9

\bibitem[Jing \& Suto(2000)]{JS00}
Jing, Y. P., \& Suto, Y. 2000, \apj, 529, L69

\bibitem[Kravtsov \& Klypin (1999)]{KK99}
Kravtsov, A. V., \& Klypin, A.  1999, \apj, 520, 437

\bibitem[Ma(1998)]{M98}
Ma, C.-P.  1998, \apj, 508, L5

\bibitem[Ma et al.(1999)]{MCBW99}
Ma, C.-P., Caldwell, R. R., Bode, P., \& Wang, L.  1999, \apj, 521, L1

\bibitem[Ma \& Fry(2000a)]{MF00a}
Ma, C.-P., \& Fry, J. N. 2000a, \apj, 531, L87

\bibitem[Ma \& Fry(2000c)]{MF00c}
Ma, C.-P., \& Fry, J. N. 2000c, \apj, in press (astro-ph/0005233)

\bibitem[McClelland \& Silk(1977)]{MS77}
McClelland, J., \& Silk, J. 1977, \apj, 217, 331

\bibitem[Mo \& White(1996)]{MW96}
Mo, H. J., \& White, S. D. M. 1996, \mnras, 282, 347

\bibitem[Mo, Jing, \& White(1997)]{MJW97}
Mo, H. J., Jing, Y. P., \& White, S. D. M. 1997, \mnras, 284, 189

\bibitem[Moore et al.(1998)]{MGQSL98}
Moore, B., Governato, F., Quinn, T., Stadel, J., \& Lake, G. 1998, 
\apj, 499, L5

\bibitem[Moore et al.(1999)]{MQGSL99}
Moore, B., Quinn, T., Governato, F., Stadel, J., \& Lake, G. 1999, 
MNRAS, 310, 1147

\bibitem[Navarro, Frenk, \& White(1996)]{NFW96}
Navarro, J. F., Frenk, C. S., \& White, S. D. M. 1996, 
\apj, 462, 563 

\bibitem[Navarro, Frenk, \& White(1997)]{NFW97}
Navarro, J. F., Frenk, C. S., \& White, S. D. M. 1997, 
\apj, 490, 493 

\bibitem[Neyman \& Scott(1952)]{NS52}
Neyman, J., \& Scott, E. L. 1952, \apj, 116, 144

\bibitem[Peacock \& Dodds(1996)]{}
Peacock, J. A., \& Dodds, S. J. 1996, MNRAS, 280, L1

\bibitem[Peebles(1974)]{P74}
Peebles, P. J. E. 1974, \aap, 32, 197

\bibitem[Peebles(1980)]{P80}
Peebles, P. J. E. 1980, The Large-Scale Structure of the Universe 
(Princeton: Princeton Univ. Press)

\bibitem[Peebles \& Groth(1975)]{PG75}
Peebles, P. J. E., \& Groth, E. J. 1975, \apj, 196, 1

\bibitem[Press \& Schechter(1974)]{PS74}
Press, W. H., \& Schechter, P. 1974, \apj, 187, 425

\bibitem[Scherrer \& Bertschinger(1991)]{SB91}
Scherrer, R. J., \& Bertschinger, E. 1991, \apj, 381, 349 

\bibitem[seljak(2000)]{S00}
Seljak, U. 2000, astro-ph/0001493

\bibitem[Sheth \& Jain(1997)]{SJ97}
Sheth, R. K., \& Jain, B. 1997, \mnras, 285, 231

\bibitem[Sheth \& Tormenn(1997)]{ST99}
Sheth, R. K., \& Tormen, G. 1999, \mnras, 308, 119

\bibitem[Tormen, Bouchet, \& White(1997)]{TBW97}
Tormen, G., Bouchet, F., \& White, S. D. M., 1997, \mnras, 286, 865

\bibitem[Valagreas(1999)]{V99}
Valageas, P. 1998, \aap, 347, 757

\bibitem[Yano \& Gouda(1999)]{YG99}
Yano, T., \& Gouda, N. 1999, astro-ph/9906375

\end{thebibliography}
\end{document}